\documentclass[12pt,preprint]{aastex}

\slugcomment{Submitted to The Astrophysical Journal}

\shorttitle{Contributions of point extragalactic sources to CMB bispectrum}
\shortauthors{Arg\"ueso et al.}

\begin{document}


\title{Contributions of point extragalactic sources \\
  to the Cosmic Microwave Background bispectrum}


\author{F. Arg\"ueso\altaffilmark{1}}
\affil{Departamento de Matem\'aticas, Universidad de Oviedo, avda. Calvo Sotelo
s/n, 33007 OVIEDO (Spain)}

\author{J. Gonz\'alez-Nuevo\altaffilmark{2} and  L. Toffolatti\altaffilmark{2}}
\affil{Departamento de F\'\i{sica}, Universidad de Oviedo, avda. Calvo Sotelo
s/n, 33007 OVIEDO (Spain)}
\email{argueso@pinon.ccu.uniovi.es, jgng@pinon.ccu.uniovi.es, toffol@pinon.ccu.uniovi.es}




\begin{abstract}
All the analyses of Cosmic Microwave Background (CMB) temperature maps
up--to--date show that CMB anisotropies follow a Gaussian distribution.
On the other hand, astrophysical foregrounds which hamper the detection of
the CMB angular power spectrum, are not Gaussian distributed on the sky.
Therefore, they should give a sizeable contribution to the CMB bispectrum.
In fact, the first year data of the Wilkinson Microwave Anisotropy Probe
(WMAP) mission have allowed the {\it first} detection of the extragalactic
source contribution to the CMB bispectrum at 41 GHz and, at the same time,
much tighter limits than before to non--Gaussian primordial fluctuations.
In view of the above and for achieving higher precision in current and future
CMB measurements of non--Gaussianity, in this paper we discuss a comprehensive
assessment of the bispectrum due to either uncorrelated or clustered
extragalactic point sources in the whole frequency interval around the
CMB intensity peak.

Our calculations, based on current cosmological evolution models for sources,
show that the reduced angular bispectrum due to point sources, $b_{ps}$,
should be detectable in all WMAP and Planck frequency channels.
We also find agreement with the results on $b_{ps}$ at 41 GHz coming from
the analysis of the first year WMAP data . Moreover, by comparing $b_{ps}$
with the primordial reduced CMB bispectrum, we find that only the peak value
of the primordial bispectrum (which appears at $l\simeq 200$) results greater
than $b_{ps}$ in a frequency window around the intensity peak of the CMB.
The amplitude of this window basically depends on the capability of the
source detection algorithms (i.e., on the achievable flux detection limit,
$S_{lim}$, for sources). Finally, our current results show that, at low
frequencies (i.e., $\nu\leq 100$ GHz), the angular bispectrum of a clustered
distribution of sources appears not substantially different from that of
Poisson distributed ones, by using realistic angular correlation functions
suitable to be applied to the relevant source populations. On the other hand,
we also find that at higher frequencies (i.e., $\nu\geq 300$ GHz),
the clustering term can greatly enhance the normalization of $b_{ps}$.

\end{abstract}




\section{INTRODUCTION}

The analysis of the first year Wilkinson Microwave Anisotropy probe (WMAP) data clearly show that temperature fluctuations of the Cosmic Microwave Background (CMB) follow a Gaussian distribution \citep{kos03,spe03} in agreement with the standard inflation paradigm.
Therefore, the third moment of this distribution and its angular bispectrum, $C_{{l_1}{l_2}{l_3}}$, the harmonic transform of the three--point correlation function, should be zero and the angular power spectrum, $C_{\ell}$, specify all the statistical properties of CMB anisotropy. Anyway, significant non-Gaussianity can be introduced, for instance, by second order relativistic effects or by features in the scalar field potential, which can produce an angular bispectrum
potentially detectable in all--sky maps as those provided by the WMAP and Planck satellite missions \citep{ben02,man98,pug98}.

In the standard inflationary scenario, the quantum fluctuations of the scalar field are the origin of the matter and radiation fluctuations in the Universe.
In slow-roll inflation, weak non-Gaussian fluctuations are generated by non-linearity in inflation \citep{sal90} or by features appearing in the inflation potential \citep{gan94} which can produce small deviations from gaussianity in the CMB temperature fluctuations.
Second order relativistic effects can give also rise to a detectable
non-Gaussian contribution \citep{pyn96}. All these effects can be summarized by the following expression for the curvature perturbations,

\begin{equation}
\Phi(x) = \Phi_{L}(x) + f_{nl}\,(\Phi_{L}^{2}(x)-\langle \Phi_{L}^{2}(x) \rangle).
\end{equation}

\noindent where $\Phi_{L}(x)$ is the Gaussian part of the perturbations, $f_{nl} $ is the non-linear coupling parameter, 
given by a certain combination of the slope and the curvature of the inflaton potential, and $\langle\rangle$ means the statistical ensemble average (Falk, Rangarajan, and Srednicki 1993; Gangui et al. 1994; Wang and Kamionkowski 2000). 
In general, high values of $f_{nl}$ can be obtained just with a small
tilt of the spectrum and these values could give rise to a bispectrum signal potentially detectable in current as well as future all sky CMB maps. 
Moreover, the value of $f_{nl}$ could be also increased by deviations from standard inflationary models (see, e.g., Kofman et al. 1991).

The great astrophysical interest in the detection of primordial non--Gaussianity has stimulated
the development of many different me\-thods suitable in analyzing CMB anisotropy: Minkowski
functionals on the sphere (Schmalzing and Gorski 1998), statistics of excursion sets (Barreiro et al. 2001), wavelets (Pando et al. 1998; Barreiro et al. 2000; Cay\'on et al. 2001), the bispectrum (Luo 1994; Komatsu and Spergel, 2001; 
Komatsu et al. 2002; Santos et al. 2002). The main result of all these analyses is that
{\it no positive} detection of non-Gaussianity in CMB maps has been confirmed up to now: COBE DMR \citep{b&t99,cay03,kos02} Maxima-1 (Wu et al. 2001; Santos et al. 2002) and
Boomerang (Polenta et al. 2001) data appear all compatible with the Gaussian hypothesis.
On the other hand, Ferreira et al. (1998) have measured 9 equilateral modes $(\ell_1=\ell_2=\ell_3)$ of the normalized
bispectrum, $B_{{l_1}{l_2}{l_3}}/C_{\ell}^{1/2}$, on the COBE DMR map, claiming detection at $\ell_1=\ell_2=\ell_3=16$ whereas \citet{b&t99} find that few individual pixels -- which could be contaminated by both foreground emission and instrumental noise -- are responsible of most of the claimed primordial signal.

A great challenge to precise measurements of non--Gaussianity in the CMB is set by astrophysical foregrounds
which constitute an unavoidable limitation : see, e.g., {\it ``Microwave Foregrounds''}, 1999, A. de Oliveira
Costa and M. Tegmark eds., ASP Conf. Ser. Vol.181, for a thorough discussion on the subject. At present, even
applying the most efficient me\-thods for component separation, the residual astrophysical signal -- which is not,
in general, Gaussian distributed -- can contaminate a non negligible number of pixels giving rise to a
detectable contribution to non--Gaussianity. Other studies have discussed the effect of Galactic foregrounds on
non--Gaussianity (see, e.g., Komatsu et al., 2002). In this paper, we focus on extragalactic point sources whose
contribution to CMB anisotropies has been analyzed in detail (Toffolatti et al. 1998; Sokasian et al., 2001;
Park, Park \& Ratra 2002) whereas their contribution to the CMB bispectrum has been poorly studied up to now. A
first estimate of the reduced bispectrum, $b_{ps}$, due to Poisson distributed point sources at two frequencies
only (90 and 217 GHz) has been recently presented by Komatsu \& Spergel (2001), hereafter KS01. This paper has
shown that undetected point sources, i.e. sources at fluxes below the detection threshold of the experiment,
give rise to a non zero $b_{ps}$ value which could hamper the detection of the primordial signal. Very recently,
by the analysis of the first year WMAP observations, \citet{kos03} have claimed the first detection of the
bispectrum due to extragalactic sources. In view of the above, it is clearly of great interest to perform a
thorough analysis of the contribution of point sources to the CMB bispectrum. Therefore, we extend the
calculation of KS01 to the whole frequency range of the WMAP and Planck missions, using model number counts able
to reproduce the data available so far. Moreover, we take also into account the effect of correlated positions
of point sources in the sky for studying how much source clustering can enhance the value of the reduced angular
bispectrum, $b_{ps}$.

The outline of the paper is as follows. In Section 2 we briefly summarize the formalism used to estimate the reduced bispectrum and show detailed calculations of its value due to Poisson distributed point extragalactic sources.
In Section 3, we model the clustering properties of sources at microwave frequencies by 2D simulations and we present the first estimates to date of the CMB bispectrum produced by
clustered sources.
Finally, in Section 4, we discuss our main results and present the conclusions.
We assume a standard Cold Dark Matter (CDM) cosmology ($\Omega_M$=1.0, $H_0=50$ km/s/Mpc) but we discuss our findings
also in the case of a flat $\Lambda$CDM model with $\Omega_M=0.3$ and $\Omega_{\Lambda}=0.7$.
Anyway, as discussed below, the conclusions are largely independent of the adopted cosmological model.




\section{THE BISPECTRUM DUE TO UNCLUSTERED POINT SOURCES}

\subsection{General formalism}

CMB temperature fluctuations ${\Delta T(\mathbf{n})}/{T}$ are usually expanded into spherical harmonics
\begin{equation}
a_{lm}=\int_{\Omega} d^{2} {\mathbf{n}}\ \frac {\Delta T(\mathbf {n})}{T} Y_{lm}^{\ast}(\mathbf{n})\,
\end{equation}
\noindent where $\Omega$ denotes the full sky (in the case of an incomplete sky coverage, the integral is
done over the observed sky area, $\Omega_{obs}$; see, e.g., Komatsu et al., 2002).
The angular third moment of CMB temperature fluctuations is defined as
\begin{equation}
B_{l_{1}l_{2}l_{3}}^{m_{1}m_{2}m_{3}}\equiv {\langle a_{l_{1}m_{1}}a_{l_{2}m_{2}}a_{l_{3}m_{3}}\rangle }
\end{equation}
\noindent and the averaging is over the ensemble of realizations.
If the fluctuations are gaussian distributed, the third moment and all the higher order moments are zero.

Due to the rotational invariance of the universe
\begin{equation}
B_{l_{1}l_{2}l_{3}}^{m_{1}m_{2}m_{3}}= \left(\begin{array}{c c c} l_{1}& l_{2}& l_{3}\\
m_{1}&m_{2}&m_{3}\end{array}\right) C_{l_{1}l_{2}l_{3}}
\end{equation}
\noindent where $ C_{l_{1}l_{2}l_{3}} $ is the angular bispectrum and the matrix is the Wigner-3j symbol.
Since we only have access to one sky we need an estimator of the angular
bispectrum: the best unbiased estimator (Gangui and Martin 2000) is the angle--averaged bispectrum
\begin{equation}
B_{l_{1}l_{2}l_{3}}\equiv\sum_{m_{1}m_{2}m_{3}}\left(\begin{array}{c c c} l_{1}&l_{2}&l_{3} \\
m_{1}& m_{2}& m_{3}\end{array}\right) B_{l_{1}l_{2}l_{3}}^{m_{1}m_{2}m_{3}}
\end{equation}

Another related quantity is the reduced bispectrum $b_{l_{1}l_{2}l_{3}}$, which can be expressed easily in
terms of the angular bispectrum

\begin{equation}
C_{l_{1}l_{2}l_{3}}= \sqrt{\frac{(2l_{1}+1)(2l_{2}+1)(2l_{3}+1)}{4\pi}}
\times \left(\begin{array}{c c c} l_{1}&l_{2}& l_{3}\\ m_{1}&m_{2}&m_{3}\end{array}\right) b_{l_{1}l_{2}l_{3}}
\end{equation}
\noindent and which contains all the physical information in $B_{l_{1}l_{2}l_{3}}^{m_{1}m_{2}m_{3}}$.
The reduced bispectrum can also be related to the third moment
\begin{equation}
 {\langle a_{l_{1}m_{1}}a_{l_{2}m_{2}}a_{l_{3}m_{3}}\rangle}={\cal G}_{l_{1}l_{2}l_{3}}^{m_{1}m_{2}m_{3}}\, b_{l_{1}l_{2}l_{3}}
\end{equation}
\noindent where ${\cal G}_{l_{1}l_{2}l_{3}}^{m_{1}m_{2}m_{3}}$ is the Gaunt integral
\begin{equation}
{\cal G}_{l_{1}l_{2}l_{3}}^{m_{1}m_{2}m_{3}}=\int d^{2} ({\mathbf{n}})
Y_{l_{1}m_{1}}({\mathbf{n}})\,Y_{l_{2}m_{2}}({\mathbf{n}})\,Y_{l_{3}m_{3}}({\mathbf{n}})
\end{equation}
\noindent The reduced angular bispectrum is a very convenient quantity, since it can
be easily calculated in the flat two-dimensional approximation (Santos et al. 2001).

Having summarized the general formalism, we remind here the calculation of the reduced bispectrum
expected in the case of unclustered point sources, i.e extragalactic point--like sources which follow
a Poisson distribution in the sky. \citet{tof98} found that this assumption is fairly
good at microwave frequencies, if sources are not subtracted down to very faint flux limits,
thus greatly decreasing Poisson fluctuations whereas leaving almost unaffected the clustering term.
Under this assumption, the sky distribution of extragalactic sources produces a white noise spectrum and,
from Eq.(7), we define
\begin{equation}
b_{ps}\equiv b_{l_{1}l_{2}l_{3}}^{sources}={\langle(T-\langle T\rangle )^{3}\rangle/T^3 }
\end{equation}
i.e., equal to the skewness of the sky temperature distribution. Therefore, $b_{ps}$=constant at all scales $\ell$.

The reduced bispectrum, $b_{ps}$, can be then calculated as follows
\begin{equation}
b_{l_{1}l_{2}l_{3}}=g^{3}(x)\int_{0}^{S_{lim}} dS \,S^{3}\,\frac{dn}{dS}
\end{equation}
\noindent where $\displaystyle\frac{dn}{dS}$ is the differential source count per unit solid angle, $S$ the flux
and $g$ the conversion factor from fluxes to temperatures
\begin{equation}
 g(x)\equiv 2\frac{(hc)^{2}}{(k_BT)^{3}}\frac{(sinh {x}/{2}))^{2}}{x^{4}}
\end{equation}
where $x\equiv h\nu/k_BT$. The integral in Eq.(10) has to be computed up to the flux detection limit
foreseen for the experiment, $S_{lim}$, since only undetected sources
are contributing to the estimated $b_{ps}$.

\subsection{Number counts of extragalactic sources}

To perform the integral in Eq. (10) we have used the differential counts corresponding to the cosmological
evolution models for radio and far--IR selected sources discussed in \citet{tof98} (hereafter TO98).
This choice represent a clearly better approximation than simple power law counts, as assumed by
KS01 with the purpose of obtaining a first estimate of $b_{ps}$. However, since they used the model counts
of TO98 for the extrapolation of source counts at 94 GHz, with the flux detection threshold of $S_{lim}\simeq 2$
Jy (the estimated $5\sigma$ detection limit of WMAP W-band), it is not surprising that their estimates result in good agreement with our current ones.

As for data on radio counts, preliminary measurements at 15.2 GHz by Taylor et al. (2001) suggested that
the extrapolation of the model counts of TO98 -- which successfully account for the observed source counts down to $S\sim 0.1$ mJy in flux and up to 8.44 GHz
--  overestimates their data by a factor of $\sim 1.5\div 2$ at the survey limit ($\simeq 20$ mJy), implying that the simple assumptions on source spectra and/or on cosmological evolution could not have been the most correct ones.
On the other hand, the recently published WMAP analysis of the foreground emission
\citep{ben03} has provided a full sky catalogue of 208 bright extragalactic sources with fluxes $S\geq 0.9-1.0$ Jy, of which only five objects could be spurious identifications.
\footnote{Only very few sources are detected by WMAP at fluxes $S< 0.9$ Jy and only two at galactic latitude
$\vert b^{II}\vert< 10^{\circ}$ according to Table 5 of \citet{ben03}. Moreover, at fluxes $S<1.2\div 1.3$ Jy the sample
appears to be statistically incomplete.}
The whole sample give an average {\it flat} ($\alpha=0.0\pm 0.2$)
energy spectrum in full agreement with the assumptions of TO98
and number counts of bright sources at 41 GHz which appear to fall below
the prediction of TO98 at 30 GHz by a factor of $\sim 0.66$
(but only in the faintest flux bins). Direct calculations of the
33 GHz counts by the source catalogue in Table 5 of \citet{ben03}
give 155 sources at $S_{lim}\simeq 1.25$ Jy on a sky area of 10.38 sr
($\vert b^{II}\vert> 10^{\circ}$) where the sample appears to be statistically
complete. By this sample we could estimate again the WMAP differential counts
finding an average offset of $\sim 0.75$ with the TO98 model predictions
(see \citet{tof03} for more details).

Moreover, two other recently published independent samples of extragalactic sources at 31 and 34 GHz -- from CBI
\citep{mas03} and VSA \citep{tay03} experiments, respectively -- show that the TO98 model {\it correctly}
predicts number counts down to, at least, $S\simeq 10$ mJy. Another independent check of the assumptions of the
TO98 model is coming from the few detections of extragalactic point sources in the Boomerang maps. These sources
show flat or decreasing spectra up to 240 GHz \citep{cob03} which is again in agreement with the population
mixture of TO98. Therefore, given that almost all up--to--date observations are confirming the predictions on
source counts of the TO98 model up to 30-34 GHz, we can be very confident in applying this model for our current
predictions, at least up to frequencies $\leq 100\div 150$ GHz, where ``flat''--spectrum sources dominate the
bright counts.

On the other hand, a lot of new data on far--IR/sub--mm source counts have piled up
since 1998, in particular by means of the SCUBA and MAMBO surveys.
These current data are better explained by new physical evolution models of far--IR selected sources, as the
ones presented by \citet{gui98} and by \citet{gra01}. Therefore, we also compare our first
estimate of $b_{ps}$ at 545 GHz with the ones obtained by directly integrating the differential counts of these two latter predictions. Anyway, since the reduced bispectrum due to undetected sources is produced by {\it} all sources below a 
given detection limit, the integral in eq. (10) results not much affected by the adopted evolution model (see below).

In Table 1 we present our current estimates of the reduced bispectrum, $b_{ps}$, due to
Poisson distributed extragalactic sources
at various frequencies in the range 15--850 GHz and applying different flux limits, $S_{lim}$, for source detection.
The frequencies and flux detection limits has been chosen to directly compare with those of the MAP and Planck satellite missions. All our estimates have been calculated by taking into account
both the radio and far--IR source populations at each frequency channel. Anyway, at frequencies
below $\sim 150$ GHz the contribution of far--IR selected sources is negligible
whereas at $\nu \geq$500 GHz the same applies to radio selected ones. The values of $b_{ps}$ are, obviously, at a minimum close to the CMB intensity peak where bright sources reach a minimum number.

It is worthy to remind that the first estimate of KS01 gave the value $b_{ps}\sim 2\times 10^{-25}$ at 94 GHz, with
$S_{lim}\sim 2$ Jy whereas our current estimate is $\sim 3.7\times 10^{-25} $; at 217 GHz and with
$S_{lim}\sim 0.2$ Jy their value was $\sim 5\times 10^{-28}$ whereas we obtain $\sim 3.8\times 10^{-28}$. The general agreement and, at the same time, the little mismatch with our
estimates are not surprising since KS01 used the TO98 model for estimating $b_{ps}$
but they extrapolated TO98 number counts with a constant slope down to faint fluxes.

\begin{table}
\begin{center}
\caption{Sample values of $b_{ps}$ in the 15--850 GHz frequency interval
and for different flux detection limits, $S_{lim}$. All the Planck channels are indicated.\label{tbl-1}}
\smallskip\smallskip
\begin{tabular}{ccccc}
\tableline\tableline
 $\nu$(GHz) & $S_{lim}=2.$ Jy & $=1.$ Jy & $=0.1$ Jy & $=0.01$ Jy \\  \hline%
\tableline
15 &$1.5\times 10^{-20}$&$5.0\times 10^{-21}$&$0.9\times 10^{-22}$&$1.0\times 10^{-24}$\\%
23\tablenotemark{a} &$7.8\times 10^{-22}$&$2.6\times 10^{-22}$&$4.6\times 10^{-24}$&$5.0\times 10^{-26}$\\%
30 &$1.9\times 10^{-22}$&$6.4\times 10^{-23}$&$1.2\times 10^{-24}$&$1.4\times 10^{-26}$\\%
44 &$2.2\times 10^{-23}$&$7.2\times 10^{-24}$&$1.3\times 10^{-25}$&$1.4\times 10^{-27}$\\%
61\tablenotemark{a} &$3.6\times 10^{-24}$&$1.1\times 10^{-24}$&$1.7\times 10^{-26}$&$1.9\times 10^{-28}$\\%
70 &$1.6\times 10^{-24}$&$5.3\times 10^{-25}$&$8.8\times 10^{-27}$&$1.0\times 10^{-28}$\\%
94\tablenotemark{a} &$3.7\times 10^{-25}$&$1.2\times 10^{-25}$&$1.9\times 10^{-27}$&$2.2\times 10^{-29}$\\%
100 &$2.8\times 10^{-25}$&$9.0\times 10^{-26}$&$1.5\times 10^{-27}$&$1.8\times 10^{-29}$\\%
143 &$4.5\times 10^{-26}$&$1.5\times 10^{-26}$&$2.8\times 10^{-28}$&$4.5\times 10^{-30}$\\%
217 &$1.6\times 10^{-26}$&$5.7\times 10^{-27}$&$1.6\times 10^{-28}$&$8.0\times 10^{-30}$\\%
353 &$1.6\times 10^{-25}$&$7.0\times 10^{-26}$&$5.5\times 10^{-27}$&$4.0\times 10^{-28}$\\%
545 &$2.0\times 10^{-22}$&$1.0\times 10^{-22}$&$7.8\times 10^{-24}$&$3.0\times 10^{-25}$\\%
857 &$1.0\times 10^{-16}$&$4.8\times 10^{-17}$&$2.6\times 10^{-18}$&$5.6\times 10^{-20}$\\%
\tableline\tableline
\end{tabular}
\tablenotetext{a}{Central frequencies of the K-, V-, and W- MAP channels.
The two remaining channels of the MAP mission, Ka- and Q-, have not been
indicated because their central frequencies, 33 and 41 GHz, are very close
to the ones of the Planck LFI instrument at 30 and 44 GHz.}
\end{center}
\end{table}

It is also interesting to compare $b_{ps}$ estimates with the reduced CMB bispectrum obtained in
non-Gaussian models with fluctuations of the inflation potential given by eq.~(1). This
quantity has also been carefully calculated by KS01. They have shown that the equilateral bispectrum $b_{lll}$ -- i.e., the reduced angular bispectrum $b_{l_{1}l_{2}l_{3}}$ when $l_1=l_2=l_3$ -- multiplied by $l^{4}$ exhibits a peak at multipole $\ell\simeq 200$.
The height of this peak is proportional to the factor $f_{nl}$, the coupling parameter.

In Figure~1 we also compare our estimates of $b_{ps}$ with the primordial CMB equilateral bispectrum, $b_{lll}$,
at its peak value ($\ell\simeq 200$) by adopting $f_{nl}=100,10,1$. As also indicated in Table~1, we explore the
whole frequency interval around the intensity peak of the CMB. Looking at Figure~1, the peak value of the
primordial $b_{lll}$ appears greater than $b_{ps}$ in a frequency window around the CMB intesity peak:
obviously, the window sets larger by increasing the value of $f_{nl}$ and by reducing the source detection
limit, $S_{lim}$. If $f_{nl}=100$ and with $S_{lim}\simeq 0.1$ Jy, the window spans from $\sim 30$ to $\sim 500$
GHz. On the other hand, if $f_{nl}=10$ and $S_{lim}\simeq 2$ Jy, the window shrinks to $100\leq\nu\leq 350$ GHz.
The line corresponding to $f_{nl}=1$ is plotted only as a reference, since the ideal experiment requires
$f_{nl}>3$, in order to obtain $S/N>1$ \citep{kos01}.

As an example, in Figure 2 we show our estimates of $b_{ps}$ at three frequencies, 30, 44 and 70 GHz, and for
$S_{lim}$=1,0.1 Jy comparing them with the primordial equilateral bispectrum, $b_{lll}$, calculated by assuming
$f_{nl}=10$. It is apparent that at this three frequencies, and, thus, for all frequencies below $\nu\simeq
70\div 80$ GHz, $b_{ps}$ has always a higher value than the primordial CMB bispectrum if $S_{lim}\geq 1$ Jy. On
the other hand, if we are able to detect and remove sources brighter than $S_{lim}\simeq 0.1$ Jy, which could be
attaineable by Planck (Vielva et al., 2003), the peak value of the primordial CMB bispectrum should be
detectable at frequencies $\nu\geq 40\div 50$ GHz, for this choice of $f_{nl}$. We checked that fixing
$f_{nl}=100$, the limits for the detectability of the primordial bispectrum move down in frequency to $\sim 50$
and $\sim 20$ GHz, if $S_{lim}=1, 0.1$ Jy, respectively. We have limited our comparisons to these values of
$f_{nl}$ since WMAP data \citep{kos03} already imply that $f_{nl}\leq 134$, at the 95\% confidence level.

\clearpage

\begin{figure}
\plotone{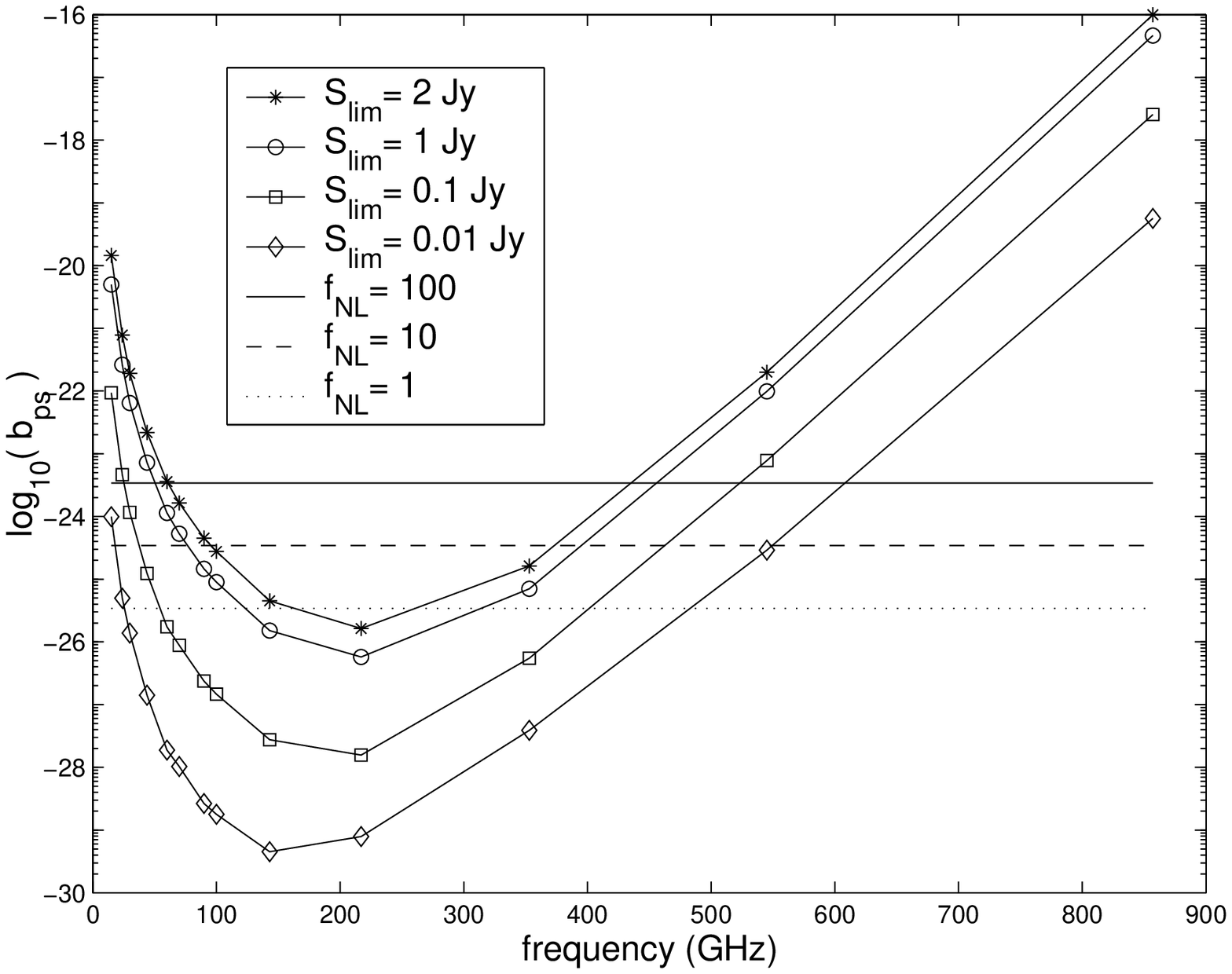}
\caption{{\small Reduced equilateral bispectrum, $b_{ps}$, due to Poisson distributed point sources
in the 15--900 GHz frequency interval and for several flux detection limits, $S_{lim}$, for sources.
The peak value of the primordial CMB bispectrum (at $\ell\sim 200$) generated by a quadratic
potential is also plotted for different values of the coupling parameter, $f_{nl}$.}}
\end{figure}

\clearpage

\begin{figure}
\plotone{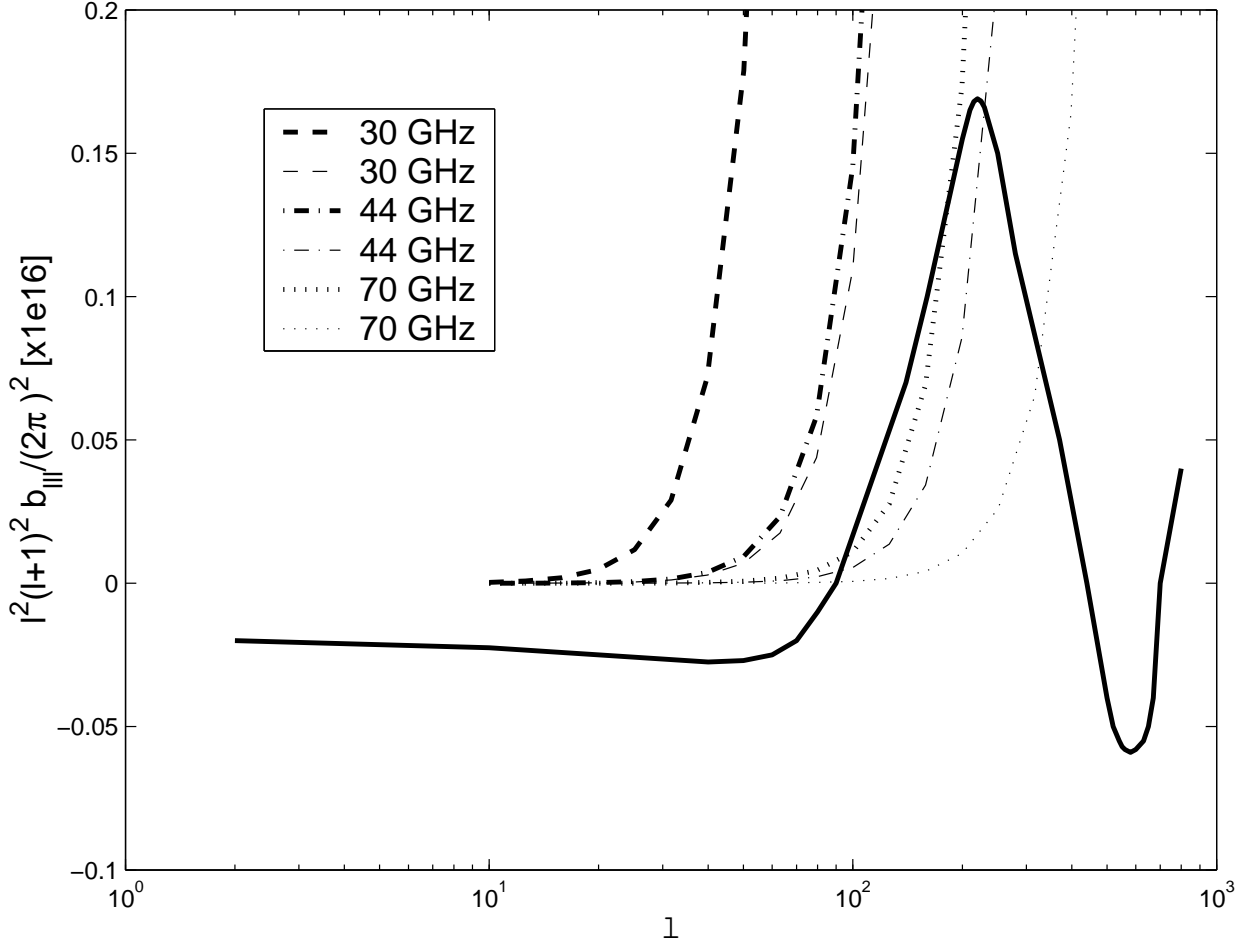}
\caption{{\small Comparison of the reduced CMB bispectrum, in terms of $\sim \ell^4 b_{lll}$, for $f_{nl}=10$, thick continuous line, with the estimates of $b_{ps}$ given by Poisson distributed sources at 30, 44 and 70 GHz and with $S_{lim}=1$ (thick dashed, point-dashed and dotted lines, respectively). We also plot the result with $S_{lim}=0.1$ Jy,
thin lines.}}
\end{figure}


As shown by KS01, the contribution of point sources and the one coming
from the Sunyaev-Zeldovich effect can be separated from the primordial CMB angular bispectrum via chi-square minimization, by taking into account the different shape of the three contributions and thanks to the sensitivity of satellite observations.
According to Table~3 in their paper, the signal-to-noise ratios for $b_{ps}$ are $2.2\times 10^{25} b_{ps}$ for WMAP and $52\times 10^{27} b_{ps}$ for Planck, respectively. The source contribution to the CMB bispectrum should therefore be detectable by WMAP if $b_{ps}\geq 2\times 10^{-25}$ (4$\sigma$ limit) and, more easily, by Planck since it should be sufficient that $b_{ps}\geq 10^{-28}$ (4$\sigma$), in this latter case.
A comparison of the above detection limit for Planck with
our current estimates of Table 1 shows that $b_{ps}$ should be detectable in {\it all} Planck channels even lowering the source detection limit down to $S_{lim}\sim 0.1$ Jy, which currently appears as an achievable limit, at least for all frequencies below $\sim 200$ GHz \citep{vie03}.
If we take the WMAP limit, we find that $b_{ps}$ could be detected also at all WMAP frequencies (only at $1\div 2\sigma$ level in the 94 GHz channel) if only sources {\it not fainter} than $S_{lim}\sim 1.0$ Jy are detected and removed. This appears as a realistic possibility given that only very few sources in Table 5 of \citet{ben03} show fluxes
$S< 0.9\div 1.0$ Jy.

Very recently, \citet{kos03} published the {\it first} detection of $b_{ps}$ at microwave frequencies: they report the value of $(9.5\pm 4.4)\times 10^{-5}$ $\mu K^3$sr$^2$,
by adopting $S_{lim}\simeq 0.75$ Jy for source detection in the 41 GHz WMAP channel.
Our current findings are in good agreement with the claim of \citet{kos03}:
in fact, by directly integrating the differential source counts of TO98 up to
$S_{lim}=0.75$ Jy at 41 GHz, we find $b_{ps}\simeq 13\div 14\times 10^{-5}$
$\mu K^3$sr$^2$, where the scatter is due to the different cosmology
($\Omega_M=1.0$ or $\Omega_M=0.3$). These results are compatible with the
detection of \citet{kos03} at $1\sigma$ level whereas we have to apply a correction
factor of $\simeq{0.7}$ for obtaining the best fit value of \citet{kos03}. If we use the more realistic
value of $S_{lim}\simeq 1.0$ Jy (see Section 2) for source detection we find 
$b_{ps}\simeq 20\div 22\times 10^{-5}$ $\mu K^3$sr$^2$ in the Q band. On the other hand, 
by integrating the TO98 model counts at 61 GHz (WMAP V band) up to $S_{lim}=0.75$ Jy we obtain
$b_{ps}\simeq 1.3\div 1.4\times 10^{-5}$ $\mu K^3$sr$^2$, showing a reduced scatter 
of $\simeq 0.80$ with the \citet{kos03} best fit value (see their Table 1). 
Moreover, \citet{hin03} have found an excess angular power spectrum
$C_{\ell}=(15.5\pm 1.7)\times 10^{-3}$ $\mu K^2$sr at small scales, which they explain as produced by undetected point sources, in agreement with the findings of \citet{kos03} . Again, by adopting $S_{lim}=$0.75 and 1.0 Jy for source detection, our current estimates at 41 GHz give $C_{\ell}\simeq$ 20 and 24.0 $\mu K^2$sr, respectively, showing a similar offset as for the point source bispectrum. Again, in the V band we obtain $C_{\ell}\simeq 4.5\div 5.0\times 10^{-3}$ $\mu K^2$sr with $S_{lim}=0.75$ in much better agreement with the \citet{kos03} 
estimate of $(4.5\pm 4)\times 10^{-3}$ $\mu K^2$sr. 
As a result, and in agreement with \citet{kos03} we actually find an
offset between their current results and our estimates of $b_{ps}$. Anyway,
\citet{kos03} results and our current estimates -- obtained by directly
integrating the TO98 differential counts -- are consistent at the $1\sigma$
level and, moreover, in the V band the offset ($\sim 0.8\div 0.9$) is smaller than in the Q band.

Finally, it is important to remind that all these
estimates have been computed by adopting a Poisson distribution
of sources in the sky and a standard CDM cosmology with $\Omega_M=1.0$.
Anyway, our current result are not very much affected by the cosmology:
in fact, if we use a flat $\Lambda$CDM model ($\Omega_{\Lambda}=0.7$),
our estimates only increases of a factor $\simeq 1.1\div 1.25$
in the 15--150 GHz frequency interval and by a slightly greater factor
($\simeq 1.5\div 2.3$) at higher frequencies, while keeping the same
parameters as in TO98 for the cosmological evolution of sources.

\section{THE BISPECTRUM DUE TO CLUSTERED POINT SOURCES}

TO98 found that the clustering contribution to CMB fluctuations due extragalactic sources is generally small in
comparison with the Poisson one. On the other hand, if clustering of sources at high redshift is very strong
it can give rise to a power spectrum stronger than the Poisson one \citep{per03,s&w99}. We remind, however, that current
models \citep{tof98,gui98} suggest a broad redshift distribution of extragalactic sources contributing to the
CMB angular power spectrum, implying a strong dilution of the clustering signal \citep{tof99}.
Anyway, in view of extracting the most precise information from current and future CMB data, i.e. pursuing {\it precision
cosmology}, it is also important to assess to which extent the
clustering signal can reduce the detectability of the primordial CMB bispectrum.

In a companion paper \citep{tof03} we have addressed the problem of estimating the CMB angular power spectrum given by undetected clustered point sources by adopting a simple {\it phenomenological} approach.
In that paper we exploit the very few data available so far on the angular correlation function of radio and far-IR selected sources to estimate the $C_{\ell}$ of extragalactic sources at microwave frequencies. We use here the same approach and the same simulated maps to estimate the clustering contribution to $b_{ps}$ as previously done for the Poisson case. We focused on WMAP and Planck LFI channels, at which extragalactic radio sources, displaying a "flat" energy spectrum
\citep{ben03}, dominate the bright counts. We also give a first estimate of $b_{ps}$ at 545 GHz (a Planck HFI channel)
and we defer to a future paper \citep{gng03} for a most comprehensive multifrequency analysis.

We briefly remind here the adopted procedure. We have carried out 100 simulations in 2D--flat patches of the sky of $12^{\circ}.8\times 12^{\circ}.8$ and $25^{\circ}.6\times 25^{\circ}.6$ deg$^2$ and with
a pixel size of $\sim$1.5 arcmin for Planck and of $\sim$6 arcmin for WMAP, since maps are created in the HEALPix format
with $nside=512$ \citep{gor98} in this latter case. A different pixel size can highly increase/decrease the pixel-by-pixel
variance in presence of a strong clustering signal, and the effect has to be taken into account. In this case, at WMAP and
LFI frequencies, the effect proves negligible in agreement with the previous findings of \citet{tof98}.

As discussed in more detail elsewhere \citep{tof03}, we first distribute point sources at random in the sky with their
total number fixed by the integral counts, $N(>S)$, of the TO98 model at each given frequency. We then calculate
the Fourier transform of the density contrast map, i.e. the map defined by $\delta(\vec{x})\equiv {N(\vec{x})-<N>
\over <N>}$, where $\langle N\rangle$ is the average number of sources per pixel, obtaining a ``white noise'' power spectrum, $P(\mathbf{k})_{Poiss}=<\vert\delta_k\vert^2>$, thus normalized to the total number of sources foreseen by the model. The next step has been to modify this normalized spectrum by  the Fourier tranform of the angular correlation function, $w(\theta )$, corresponding to the appropriate source population. We adopt here the $w(\theta)$ measured at 4.85 GHz by \citet{lwl97}, since the source population of bright sources ($S>50$ mJy) which has been sampled at this frequency is representative also of bright sources seen at higher frequencies: the underlying parent population appears to be the same one \citep{ben03} in
agreement with the assumption of the TO98 model. Other estimates of $w(\theta)$ for radio
sources come from samples at lower frequencies (e.g., 1.4 GHz) and down to lower fluxes
where the dominant source populations are different from the ones relevant for WMAP and
Planck observations.

With this choice of $w(\theta)$ the modified density contrast of each pixel
in the Fourier space is
\begin{equation}
\delta_k(corr)=\delta_k\sqrt{P(\mathbf{k})_{Poiss}+P(\mathbf{k})_{cl}\over P(\mathbf{k})_{Poiss}}
\end{equation}
where $P(\mathbf{k})_{cl}$ is the Fourier transform of the chosen $w(\theta)$. In this way the source density is modified
by the adopted correlation function whereas the normalization to the total effective number of sources predicted
by the model remains fixed by $\langle N\rangle$. By the 2D Fourier antitransform we get again the map in the real space (the sky patch). As discussed in detail by \citet{tof03}, the procedure is ``safe'' since the total number of sources and the number counts remain unchanged whereas the recovered $w(\theta)$ matches very well the input one. We want to stress here that this is a {\it phenomenological} approach, since we are currently interested in reproducing the
observational angular correlation functions of the relevant source populations, if they
are available.

\clearpage

\begin{figure}
\plotone{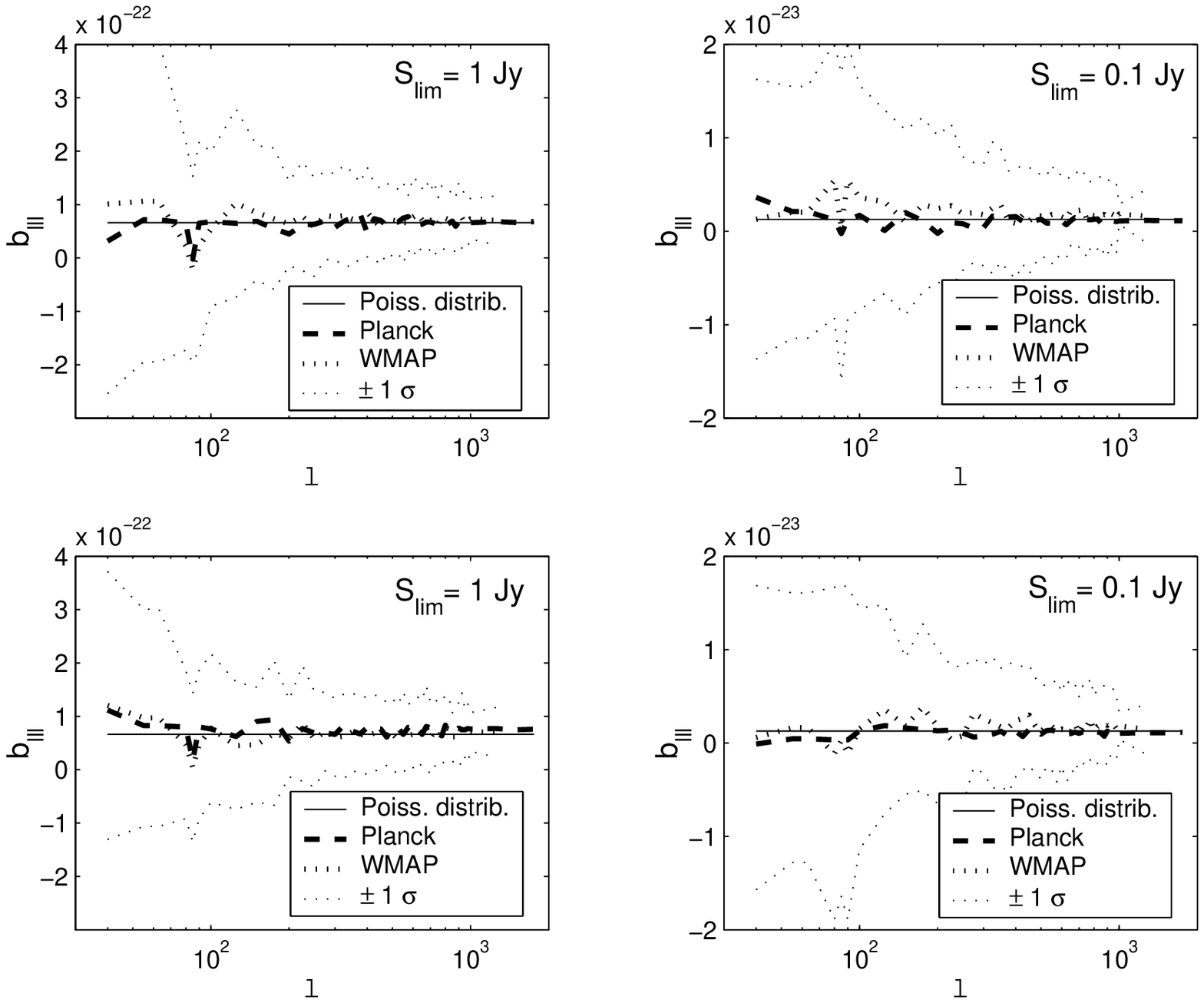}
\caption{{\small Reduced equilateral bispectrum, $b_{ps}$, at 30 GHz for two different source detection limits.
In each panel, the straight horizontal line represents the $b_{ps}$ value of Table 1.,
obtained by directly integrating Eq.(10) and, thus, using a Poisson distribution of sources in the sky. For comparison, we also plot the $b_{ps}$ values
obtained by averaging over 100 simulations of clustered point sources, upper panels, and of Poisson distributed sources, lower panels, at the Planck and WMAP resolutions (see text). In the case of WMAP we actually performed the simulations at
33 GHz. The $\pm 1\sigma$ limits refer to the set of simulations at the WMAP resolution limit.}}
\end{figure}

\noindent
Finally, from the simulated maps, in terms of $\Delta T/T$, we then obtain the Fourier transform
\begin{equation}
a({\mathbf{k}})=\int \frac {\Delta T(\mathbf {r})}{T} e^{-i{\mathbf{kr}}} \, d^{2} {\mathbf{r}}
\end{equation}
and the calculation of the bispectrum in the 2D flat-sky approximation is given by
\begin{equation}
<a({\mathbf{k_{1}}})a({\mathbf{k_{2}}})a({\mathbf{k_{3}}})>=
(2\pi)^{2}\delta^{(2)}(\mathbf{k_{1}}+\mathbf{k_{2}}+\mathbf{k_{3}})\,b(\mathbf{k_{1}},\mathbf{k_{2}},\mathbf{k_{3}})
\end{equation}
\noindent where $a({\mathbf{k}})$ is the Fourier transform of the temperature map, $\delta^{(2)}$ is the 2D Dirac delta
function and $b(\mathbf{k_{1}},\mathbf{k_{2}},\mathbf{k_{3}})$ is the reduced bispectrum.

This formula is given by the collapse of the Gaunt integral, Eq.(8), to the Dirac delta function in the flat approximation. We then estimate the reduced bispectrum by means of the following estimator, similar to the one used
by \citet{san02},

\begin{equation}
b_{l_{1}l_{2}l_{3}}=\frac{1}{N_{l_{1}l_{2}l_{3}}}\sum
\textit{Re}[a({\mathbf{k_{1}}})a({\mathbf{k_{2}}})a({\mathbf{k_{3}}})]
\end{equation}
where $\textit{Re}$ means the real part (which guarantees that the estimator is real).
Finally, we average over the combination of modes with multipoles $l_{1},l_{2},l_{3}$,
which satisfies the condition ${\mathbf{k_{1}}}+{\mathbf{k_{2}}}+{\mathbf{k_{3}}}=0$, taking into account Eq.(14).
The plotted values refer always to the reduced equilateral bispectrum, i.e. for $l_{1}=l_{2}=l_{3}$.

In Figure 3 we plot the reduced equilateral bispectrum, $b_{ps}$, calculated at 30 GHz and with source
detection limits of $S_{lim}=$1,0.1 Jy. As discussed before, we calculate the average values over 100 simulations and
the corresponding $\pm 1\sigma$ confidence intervals. To avoid the overlapping of too many lines we decided
to only plot the confidence intervals corresponding to the WMAP resolution given that the ones calculated at the Planck resolution are very close to the former ones.
From the comparison of the top panels (correlated distributions) with the bottom ones (Poisson distribution), and
with the direct estimates of $b_{ps}$, by integrating the differential counts (Eq.(10)), we can extract the following conclusions: a) from the simulations of Poisson distributed sources we can estimate $b_{ps}$ values with a very high precision. The scatter around the straight horizontal line is always much smaller than the $\pm1\sigma$ limit. b) If we
perform simulations with a correlated distribution of sources, e.g. by the $w(\theta)$ of \citet{lwl97}, the value
of $b_{ps}$ remains practically unchanged. This also happens if we choose another angular correlation function
among realistic ones which could be suitable for the underlying source population. As previously discussed,
this is due to the very broad redshift distribution of sources contributing to the bright counts in this frequency range.
On the other hand, the standard deviation results higher for the correlated distribution if compared to the Poisson case and, in particular, at low $\ell$.
In Figures 4 and 5 we plot our results on $b_{ps}$ at WMAP and Planck LFI frequencies. It is apparent that there are very small variations in comparison with the values of Table 1.

\clearpage

\begin{figure}
\plotone{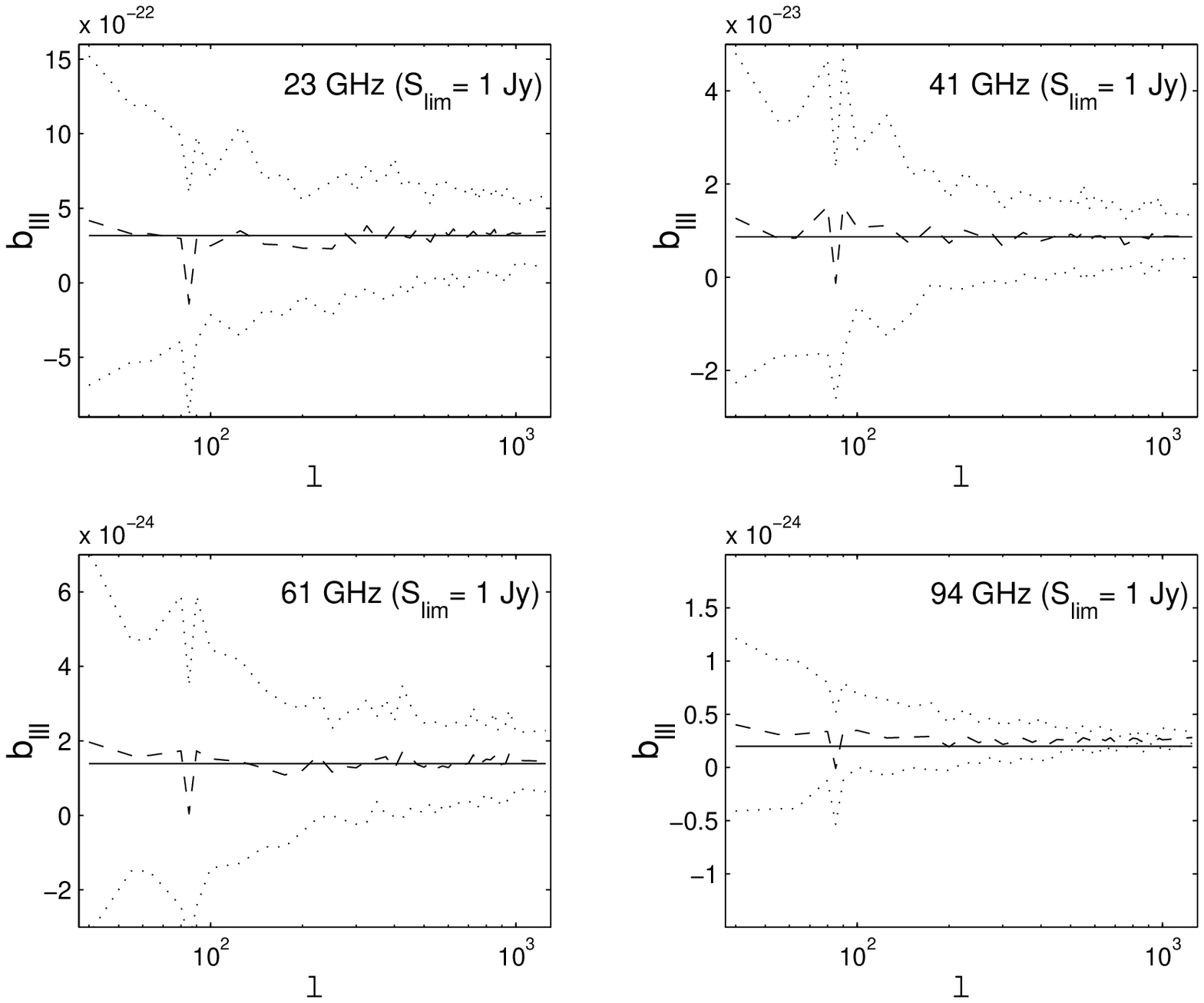}
  \caption{\small $b_{ps}$ due to a clustered distribution of point sources at four WMAP frequencies:
23, 41, 61 and 94 GHz. As in Figure 3, the straight horizontal line represent the $b_{ps}$ values of Table 1.
The dashed line represents the average $b_{ps}$ over 100 simulations and the dotted lines represent the $\pm 1\sigma$
limits.}
\end{figure}

\clearpage

\begin{figure}
\plotone{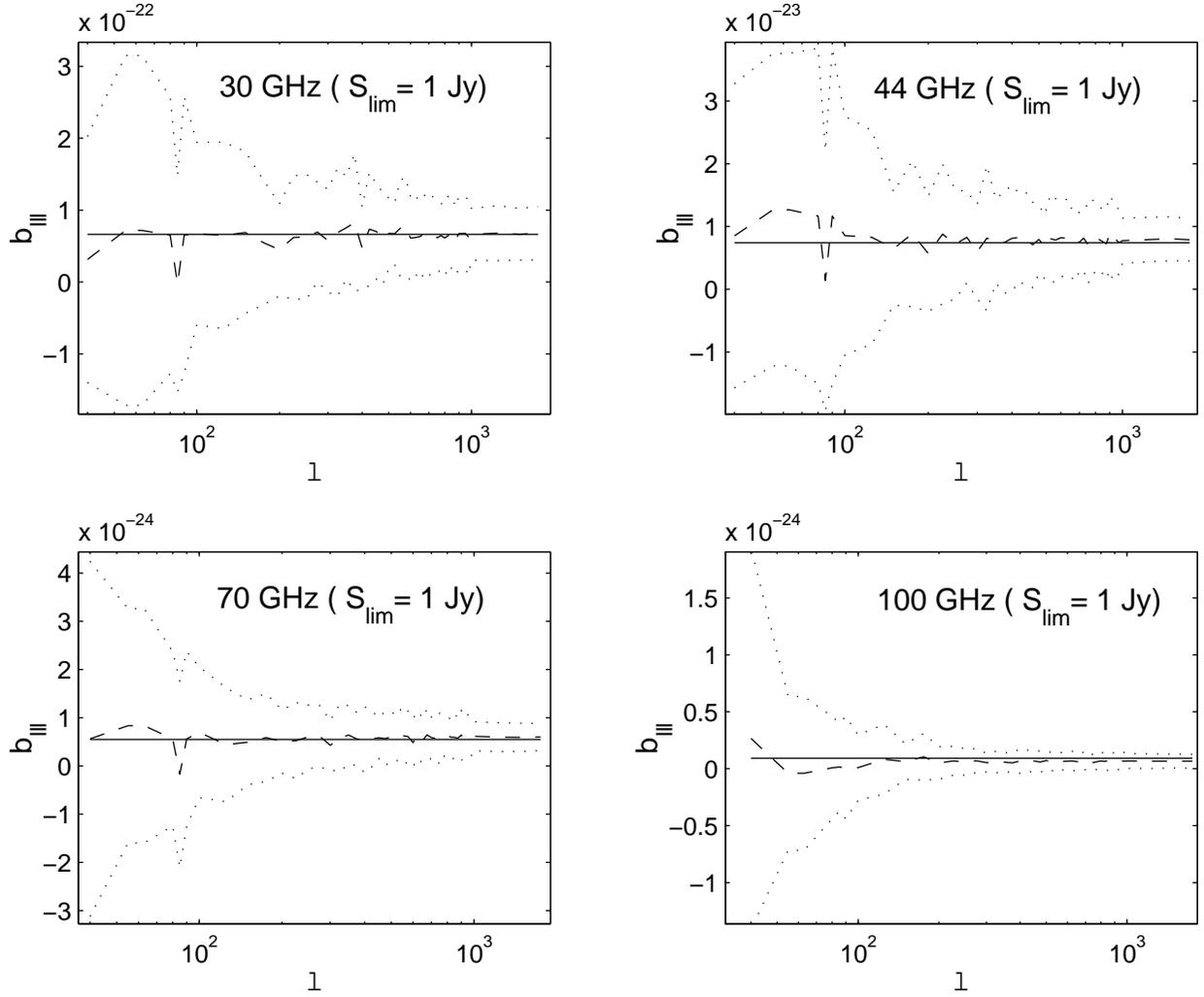}
  \caption{\small $b_{ps}$ due to a clustered distribution of point sources at Planck LFI frequencies:
30, 44, 70 and 100 GHz. The meaning of the lines is the same as in Figure 4.}
\end{figure}

On the other hand, to give a preliminary estimate of the $b_{ps}$ at Planck HFI frequencies, we have also calculated the reduced equilateral bispectrum of a non Poisson distribution of sources at 545 GHz. At this frequency the source populations dominating the number
counts are dust--enshrouded elliptical galaxies and spheroids at substantial redshift \citep{tof98,gui98,gra01} and, consequently, the clustering term should not be negligible \citep{mag01,per03}, given that the dilution of the signal is less effective than
that of sources showing a broad redshift distribution.
The results plotted in Figure 6 are obtained, as before, by distributing sources in the sky using a Poisson distribution which is then modified by the angular correlation functions appropiate for each source populations. As at lower frequencies, the value of $b_{ps}$ estimated by Eq.(15), applied to the simulated maps and with a Poisson distribution of sources, still gives a very good approximation of the value directly obtained by Eq.(10).

As for the correlated distributions, for radio selected "flat"--spectrum sources, which are giving a quite small but not negligible contribution to the bispectrum at these frequencies, we have still applied the $w(\theta)$ of \citet{lwl97}, under the assumption that the underlying parent population is the same (flat--spectrum QSOs, blazars) up to these frequencies.
As for sources whose emission is dominated by cold dust, the lack of direct data at 545 GHz forced us to rely on SCUBA as well as on optical and near--IR surveys (see \citet{per03}).
By distinguishing the relevant populations in starburts/spirals at low redshift and
ellipticals and spheroids at intermediate to high redshift, we then applied to them the $w(\theta)$ of Tegmark et al. (2002) and of \citet{mag01}, respectively. This
is a very preliminary estimate and, naturally, it will be improved in the next future. 
Anyway, given that the most relevant quantity is the total number of sources below the detection threshold, the choice of a different -- but realistic -- angular
correlation function does not modify very much the estimated $b_{ps}$: Figure 6 clearly
show that the three different evolution models for sources are giving basically the
same estimated $b_{ps}$. On the other hand, it is also apparent that at frequencies
$\geq 300$ GHz source clustering can greatly enhance the value of $b_{ps}$ and, particularly,
at multipoles $\ell \leq$100. This last result is again in agreement with
the results of \citet{s&w99,mag01,tof03} on the angular power spectrum.

\clearpage

\begin{figure}
\plotone{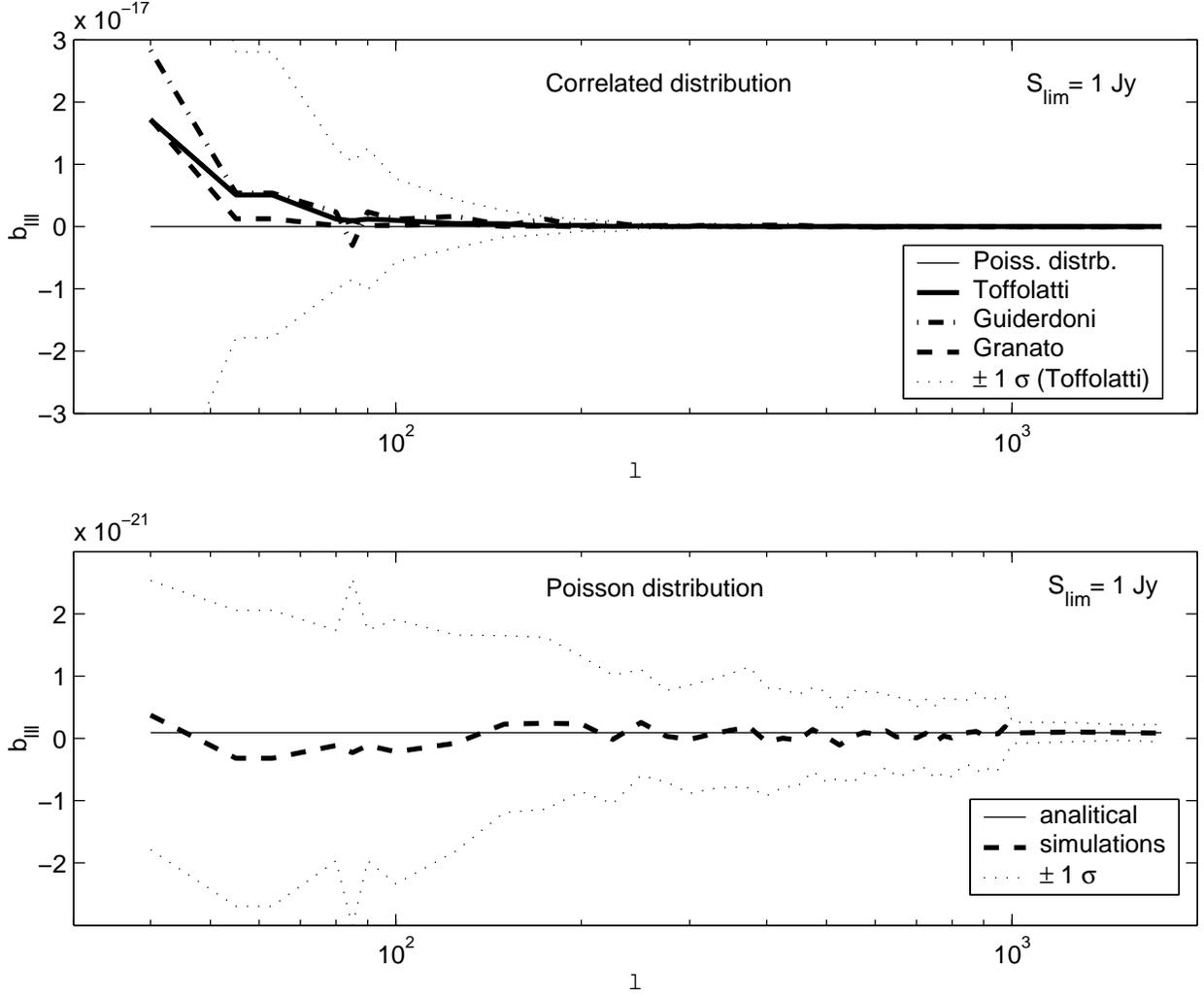}
  \caption{\small $b_{ps}$ at 545 GHz and applying the conservative source detection limit $S_{lim}=1$ Jy (see text). As in the previous Figures, the straight horizontal line represents the value of Table 1, obtained by Eq.(10): please note that in the two panels the plotted value is the {\it same one}, i.e., $b_{ps}=5\times 10^{-23}$, in spite of the great scale difference.
The top panel shows the comparison of the average values of $b_{ps}$ obtained by performing
100 simulations with the number counts of three different evolution models (see label on the
Figure). We adopted different clustered distribution of sources for each one of the three different models of sources counts (see text). The lower panel displays the comparison of the analytical estimate of $b_{ps}$ with the result of 100 simulations obtained by a simple
Poisson distribution of sources (dashed line, TO98 model counts). In the two panels, the dotted lines represent the $\pm 1\sigma$ limits obtained with the TO98 source counts.}
\end{figure}

\section{DISCUSSION AND CONCLUSIONS}

The primordial bispectrum is a telling quantity of the non-Gaussianity level of the CMB and its detection should be of great importance for theories of the early Universe. On the other hand, undetected extragalactic sources, which are not Gaussian distributed in the sky are surely giving rise to a bispectrum detectable by current as well as future CMB experiments.
In fact, \citet{kos03} already claimed the first detection of the bispectrum due to undetected point sourecs at 41 GHz by the analysis of the first year WMAP data. Therefore, the possible
detection of the primordial CMB bispectrum is severely hampered by the presence of the
foreground emission of point sources which has to be carefully evaluated.

Assuming that the sources are Poisson distributed in the sky their reduced bispectrum
is easy to calculate by integrating the differential counts of the relevant source populations at a given frequency. In this paper, we have carried out a detailed analysis of the bispectrum due to extragalactic point sources in all Planck and WMAP frequency channels.
We used the \citet{tof98} evolution model for radio and far--IR selected sources as
a template to estimate the $b_{ps}$ foreseen in the case of a Poisson distribution of point sources and by using two different cosmologies (see Section 2).
As proved by KS01, the point source bispectrum can be separated from the primordial one taking into account the difference in shape. Our current est imates proves that $b_{ps}$ can be
detected in all WMAP channels if the detection limit for sources is $S_{lim}\geq$1 Jy -- albeit at the $1\div2\sigma$ 
level at 94 GHz -- and also in all Planck channels if $S_{lim}\geq$0.1 Jy.
Table~1 and Fig.~1, 2 show in detail how the values of $b_{ps}$ of Poisson distributed sources can affect the detection of the primordial bispectrum, $b_{lll}$, in the whole frequency interval around the CMB intensity peak and at Planck LFI frequencies, respectively.

In the case of Poisson distributed sources, our main results can be summarized as follows:

a) the bispectrum detected by \citet{kos03} in the WMAP 1-year sky maps (Q and V bands)
is compatible at the $1\sigma$ level with the predictions on $b_{ps}$ calculated
by the number counts of TO98. The best fit value of $b_{ps}$ measured by \citet{kos03} in the Q band
can be explained by undetected extragalactic sources according to the TO98 model predictions
multiplied by $\simeq 0.70$, if we integrate the number counts up to
$S_{lim}=$0.75 Jy like in \citet{kos03}. On the other hand, in the V band we
find a smaller mismatch ($\simeq 0.8\div 0.9$) with current WMAP
measurements. Both uncertainties in the measured $b_{ps}$ and errors in TO98 predictions could explain the detected offsets. 
In particular, the different
correction factors which are currently found for the Q and V bands, if confirmed by future
analyses, could be indicative that the adopted spectral index distribution of sources 
has to be partially corrected. Current surveys at 15 and 31 GHz \citep{wal03,
ben03} are showing a greater fraction of "steep"--spectrum radio sources 
at intermediate to bright fluxes than that foreseen by the original TO98 model.
These new results can help in reducing the observed offsets between current
measurements of $b_{ps}$ and model predictions while keeping a good fit to
the CBI, VSA and WMAP differential source counts. Another, albeit small,
source of uncertainty is the underlying cosmology: correction factors of 10-20\% in the estimated
$b_{ps}$ values are easily introduced in this frequency range by changing $\Omega_M$ and $\Omega_{\Lambda}$.

b) The primordial CMB bispectrum, $b_{lll}$, appears higher than $b_{ps}$ only
at multipoles close to $\ell\sim 200$ where it reaches its peak value (see Figures 1 and 2).
The frequency window inside which the peak value of $b_{lll}$ is detectable sets larger by increasing $f_{nl}$ and by reducing the source detection limit, $S_{lim}$, i.e. it strongly depends on the efficiency of the source detection algorithm. If $f_{nl}=100$ and with $S_{lim}\simeq 0.1$ Jy, the window spans from $\sim 30$ to $\sim 500$ GHz. On the other hand, if $f_{nl}=10$ and $S_{lim}\simeq 2$ Jy, the window shrinks to $100\leq\nu\leq 350$ GHz. These results clearly show that $b_{ps}$ does not hamper the possible detection by Planck of very low values of the coupling parameter ($f_{nl}< 10$), comparable to the theoretical limit achievable by the mission.

c) If the detection of $b_{ps}$ is effectively achieved (e.g., by WMAP data)
the simple comparison with the $b_{ps}$ levels estimated by the integration of the differential counts of sources, $dN/dS$, allow, in principle, to test the adopted evolution model for sources. Anyway, as already noted by KS01, this can be applied 
only to the bright end of the number counts, at least for frequencies below 150$\div$200 GHz, given the `flat' slope of the counts -- close to or $\beta=2.5$, i.e.  `euclidean' -- which implies that faint fluxes are not relevant in Eq.(10).
As also discussed in Section 2., the adopted cosmology does not affect all the above conclusions.

For estimating to which extent the clustering of sources can affect the value of $b_{ps}$ we have carried out flat 2D simulations in sky patches with pixel sizes of $1.5$ (Planck) and $6$ (WMAP) arcmin.
In this analysis we simulated sources with a particular choice of the angular correlation function, but we neglected higher order moments of the distribution. This simplified approach could affect, in principle, the conclusion that source clustering does not modify the bispectrum at frequencies $\leq 100$ GHz (see Figures 3, 4 and 5). However, as discussed in Section 2 and 3, the results are much more determined by the number of undetected sources in the sky (the differential counts of bright sources) than by the choice of the correlation function or by the inclusion of higher order moments in the simulations. We checked that if we apply a different $w(\theta)$ to the same number counts, i.e. to the same evolution model for counts, the estimated $b_{ps}$ keeps practically unchanged.

On the other hand, at frequencies $\nu\geq$300 GHz, the number of
sources in the sky at a given flux limit is greatly enhanced by the rising energy spectra
due to the emission of the cold dust components. Correspondingly, if
sources do cluster down to very low flux limits, the clustering term can dominate the
Poisson one and $b_{ps}$ values result greatly enhanced (Figure 6). Therefore, it is
surely of great astrophysical interest to study the clustering properties as well
as the cosmological evolution of extragalactic sources in this frequency range for obtaining
a better assessment of their contribution to the CMB angular power spectrum and bispectrum.


\acknowledgments

We wish to thank an anonymous referee for his comments and suggestions.
We acknow\-ledge partial financial support form the Spanish MCYT under projects
ESP2001--4542--PE and ESP2002--04141--C03--01.
FA and LT also wish to thank the EC Research Training Network
contract n. HPRN--CT--2000--00124 for partial financial support.
JGN acknowledges a FPU fellowship of the Spanish Ministry of Education (MEC).
We thank B. Guiderdoni and G.L. Granato for providing us with their model
counts of sources at 545 GHz. We also acknowledge very fruitful discussions with J.L. Sanz,
E. Mart\'\i{nez}-Gonz\'alez, L. Cay\'on, R.B. Barreiro and P. Vielva.

\end{document}